

\input phyzzx

\catcode`\@=11 
\def\papers{\papersize\headline=\paperheadline\footline=\paperfootline}
\def\papersize{\hsize=40pc \vsize=53pc \hoffset=0pc \voffset=1pc
   \advance\hoffset by\HOFFSET \advance\voffset by\VOFFSET
   \pagebottomfiller=0pc
   \skip\footins=\bigskipamount \normalspace }
\catcode`\@=12 
\papers

\def\half{\textstyle{1\over 2}}
\def\cm{c_M}
\def\cl{c_L}
\def\pl{p_L}
\def\qm{Q_M}
\def\ql{Q_L}
\def\fl{{\cal F}(\pl )}
\def\fm{{\cal F}(p_M )}
\def\d{\partial}

\tolerance=500000
\overfullrule=0pt

\pubnum={PUPT-1306}
\date={February 1992}
\pubtype={}
\titlepage
\title{ REMARKS ON THE BRST-COHOMOLOGY \break
 FOR $\cm >1$ MATTER COUPLED\break
TO ``LIOUVILLE
GRAVITY"}
\author{{
Adel~Bilal}\foot{
 on leave of absence from
Laboratoire de Physique Th\'eorique de l'Ecole
Normale Sup\'erieure,
\nextline
24 rue Lhomond, 75231
Paris Cedex 05, France
(unit\'e propre du CNRS)}}
\address{\it Joseph Henry Laboratories\break
Princeton University\break
Princeton, NJ
08544, USA}

\singlespace
\abstract{We describe the
(chiral)
  BRST-cohomology of matter with
central charge $1<\cm<25$ coupled to a ``Liouville"
theory, realized as a free field with a background charge
$\ql$ such that $\cm + \cl =26$. We consider two
cases:\nextline a) matter is realized by one free field
with an imaginary background charge,\nextline
b) matter is realized by $D$ free fields: $\cm =D$.

In case a) the cohomology states can be labelled by
integers $r,s$ of a rotated $\cm =1$ theory, but
hermiticity imposes $r=s$. Thus there is still a discrete
set of momenta $p_M(r,r),\ p_L(r,r)$ such that there are
non-trivial (relative) cohomology states at level $r^2$
with ghost-numbers 0 or 1 (for $r>0$) and ghost-numbers 0
or $-1$ (for $r<0$). The (chiral) ground ring is isomorphic
to a subring of the $\cm =1$ theory which is $(xy)^n,\
n=0,1,2,\ldots$, and there are {\it no} non-trivial
currents acting on the ground ring.

In case b) there is no non-trivial relative cohomology for
non-zero ghost numbers and, for zero ghost number, the
cohomology groups are isomorphic to a $(D-1)$-dimensional
on-shell ``transverse" Fock space. The only exceptions
are at level 1 for vanishing matter momentum and $\pl =\ql
(1+r)$ with $r=\pm 1$, where one has one more ghost-number
zero and a ghost-number $r$ cohomology state.

All these results follow quite easily from the existing
literature.

\endpage
\pagenumber=1
\normalspace

 \def\PL #1 #2 #3 {Phys.~Lett.~{\bf #1} (#2) #3}
 \def\NP #1 #2 #3 {Nucl.~Phys.~{\bf #1} (#2) #3}
 \def\PR #1 #2 #3 {Phys.~Rev.~{\bf #1} (#2) #3}
 \def\PRL #1 #2 #3 {Phys.~Rev.~Lett.~{\bf #1} (#2) #3}
 \def\CMP #1 #2 #3 {Comm.~Math.~Phys.~{\bf #1} (#2) #3}
 \def\IJMP #1 #2 #3 {Int.~J.~Mod.~Phys.~{\bf #1} (#2) #3}
 \def\JETP #1 #2 #3 {Sov.~Phys.~JETP.~{\bf #1} (#2) #3}
 \def\PRS #1 #2 #3 {Proc.~Roy.~Soc.~{\bf #1} (#2) #3}
 \def\IM #1 #2 #3 {Inv.~Math.~{\bf #1} (#2) #3}
 \def\JFA #1 #2 #3 {J.~Funkt.~Anal.~{\bf #1} (#2) #3}
 \def\LMP #1 #2 #3 {Lett.~Math.~Phys.~{\bf #1} (#2) #3}
 \def\IJMP #1 #2 #3 {Int.~J.~Mod.~Phys.~{\bf #1} (#2) #3}
 \def\FAA #1 #2 #3 {Funct.~Anal.~Appl.~{\bf #1} (#2) #3}
 \def\AP #1 #2 #3 {Ann.~Phys.~{\bf #1} (#2) #3}
 \def\MPL #1 #2 #3 {Mod.~Phys.~Lett.~{\bf #1} (#2) #3}

{ \chapter{Introduction}}

Whereas the BRST-cohomology for critical strings has been
computed already some time ago
\REF\KO{M. Kato and K. Ogawa, \NP B212 1983 443 .}
\REF\FO{M.D. Freeman and D.I. Olive, \PL B175 1986 151 .}
\REF\HEN{M. Henneaux, \PL B177 1986 35 .}
\REF\FGZ{I.B. Frenkel, H. Garland and G.J. Zuckerman,
Proc. Natl. Acad. Sci. USA {\bf 83} (1986) 8442.}
[\KO-\FGZ], it is only quite recently that it has been
worked out for non-critical strings with matter central
charge $\cm \le 1$
\REF\BMP{P. Bouwknegt, J. McCarthy and K. Pilch, CERN
preprint CERN-TH.6162/91 (July 1991).}
\REF\LZ{B.H. Lian and G.J. Zuckerman, \PL B254 1991 417 ,
{\bf B266} (1991) 21.}
[\BMP,\LZ]. The basic assumption is that 2D gravity coupled
to $\cm \le 1$ matter can be described  in the conformal
gauge by coupling the matter theory to a free Liouville
field with imaginary background charge. A very
interesting structure has emerged where cohomology states
with non-zero ghost-numbers lead to a ground ring and
those with zero ghost-numbers to symmetries of that
ground ring
\REF\WIT{E. Witten, IAS preprint IASSNS-HEP-91/51 (August
1991).} [\WIT]. This structure has been used to obtain
non-trivial information about correlations functions
\REF\KMS{D. Kutasov, E. Martinec and N. Seiberg,
Princeton/Rutgers preprint PUPT-1293/RU-91-49 (November
1991); \nextline
I.R. Klebanov, Princeton preprint PUPT-1302 (December
1991).} [\KMS].

 This analysis was restricted to $\cm \le 1$.
It is very interesting to know what happens if $1<\cm
<25$. In fact, the results follow quite easily from the
analysis of refs. \FGZ, \BMP\ and  \WIT, but since we are
not aware in the literature of any explicit discussion of
this BRST-cohomology for $1<\cm <25$, we found it
nevertheless worthwhile to spell out the results in this
note. We will consider two cases:

a) The matter sector is realized by one free field with an
imaginary background charge $\qm$: $\cm=1-12\qm^2 >1$.

b) The matter sector is realized by $D$ free fields (with
euclidean or minkowskian signature): $\cm =D$.

Of course we could also consider the more general case of
$D$ free fields, one of which has a background charge.
However, by a rotation in field space we may reduce this
again to case b), and we will comment briefly on this
situation after analysing case b).

The results we find are those announced in the abstract.
Note that we always refer to the relative cohomology
$H^n_{\rm rel}$
(for one chiral sector\foot{
The non-chiral, i.e.
closed string cohomology involves some subleties
\REF\WZ{E. Witten and B. Zwiebach, IAS preprint
IASSNS-HEP-92/4 (January 1992).} [\WZ], and we will not
discuss it here
explicitly.
}),
i.e. we only consider states annihilated
by the anti-ghost zero-mode $b_0$. The absolute
cohomology then is simply $H^n_{\rm abs} \simeq
H^n_{\rm rel} \oplus c_0 H^n_{\rm rel}$.

In a certain sense, these results are quite disappointing,
since one might have hoped that some particular
BRST-cohomolgy might occur in the special dimensions put
forward in refs.
\REF\BGN{J.-L. Gervais and A. Neveu, \PL B151 1985 271 ;
\nextline
A. Bilal and J.-L. Gervais, \NP B284 1987 397 ;
\nextline
J.-L. Gervais, \PL B243 1990 85 , \CMP 138 1991 301 .}
\BGN. Although the approach of refs. \BGN\ advocates the
use of two sets of oscillators in the same Fock space of
the Liouville sector, related by a complicated
transformation, (instead of using screening operators),
the Fock spaces used (for each $\pl$) are the same as in
our analysis. Thus our results are applicable and we
conclude that, as far as only the BRST-cohomology is
concerned, nothing special happens at the special
dimensions of refs. \BGN.

In the present note, we will try as much as possible to
remain within the Fock space approach of ref. \BMP\ which
we believe to be the most physical one (although some
arguments could be short-cut by using the results of ref.
\FGZ ).

{ \chapter{ Matter is one free field with an
imaginary background charge $\qm$ (case a).}}

Formally, the analysis of the BRST-cohomology proceeds
exactly as in ref. \BMP, henceforth referred to as
BMP,\foot{
We use the same notations and conventions as BMP.} for the
product of the matter and Liouville Fock spaces $\fm \otimes
\fl$, but now the matter background charge is imaginary,
whereas in BMP it was real. However, as far as the
BRST-cohomology is concerned, the whole analysis is
insensitive to the actual value of the background charges
$\qm, \ql$ (only constrained by $\qm^2 + \ql^2=-2$ to ensure
$\cm+\cl=26$). As a result, we know that non-trivial
cohomology states appear at level $rs$ ($\ge 0$) with $r,s$
non-zero integers, iff $\pl=\pl(r,s),\ p_M=p_M(r,s)$ with
$$\eqalign{\pl(r,s)-\ql&=\ql {r+s\over 2}-i\qm {r-s\over
2}\cr
p_M(r,s)-\qm&=\qm {r+s\over 2}+i\ql {r-s\over
2}\ .}
\eqn\i$$
 Then
there is always one (relative) cohomology state with zero
ghost-number and one with ghost-number sign($r$)=sign($s$).
(For $r$ or $s$ zero, there is only a ghost-number zero
cohomology state: the corresponding Fock space vacuum.)
For $\cm\le 1$, $\qm$ was real and $\ql$ imaginary, hence
$p_M(r,s)$ was real and  $\pl(r,s)$ imaginary, which was
appropriate for hermiticity. Now, for $1<\cm<25$, we have
$\qm$ and $\ql$ imaginary and for general $r,s$ we get
complex $p_M(r,s)$ and $\pl(r,s)$. This implies that, in
general for $\pl=\pl(r,s),\ p_M=p_M(r,s)$,  the matter and
Liouville stress tensors are not hermitian separately,
only their sum is. We believe, however, that it is a
reasonable physical requirement to impose hermiticity
separately on the matter stress tensor $T_M$ and on the
Liouville one $T_L$. Let us briefly discuss the
consequences of this requirement.

Consider one free field with
background charge $Q$ and Virasoro generators
$$L_n=\half \sum_{m\ne 0,n} :a_m a_{n-m}: + \left(
p-(n+1)Q\right) a_n
\eqn\ii$$
where $p\equiv a_0$. Since the dual of the Fock space
${\cal F}(p)$ is isomorphic to ${\cal F}(2Q-p)$,
hermiticity means
$$L_n(p)^\dagger = L_{-n}(2Q-p)\ .
\eqn\iii$$
Comparing the quadratic terms simply implies $a_n^\dagger
=\pm a_{-n}$ while comparing the linear terms implies the
following:
$$\eqalign{Q\ {\rm real}\ &\Rightarrow\  a_n^\dagger
= -a_{-n}\ ,\ \ p^*=p\cr
Q\ {\rm imaginary}\ &\Rightarrow\  a_n^\dagger = +a_{-n}\
,\ \ p^*=-p\ .}
\eqn\iv$$
We see that hermiticity requires that $p$ is real if
$Q$ is real and $p$ is imaginary if $Q$ is imaginary.

Going back to our case of interest, $\qm$ and $\ql$
imaginary, we conclude that $p_M$ and $\pl$ should be
imaginary. From eq. \i\ we see that
$p_M(r,s)$  and
$\pl(r,s)$ are imaginary only for $r=s$:
$${\rm hermiticity}\  \Rightarrow\ r=s\ .
\eqn\v$$
Thus, in the hermitian case, we have non-trivial
cohomology iff
$$\eqalign{\pl=\pl(r,r)&=\ql (r+1)\cr
p_M =p_M(r,r)&=\qm (r+1)\ .}
\eqn\vi$$
In these cases the cohomology states (with ghost-numbers 0
and 1 for $r>0$ and 0 and $-1$ for $r<0$, and 0 for $r=0$)
are formally (i.e. when expressed in terms of $\qm,\ql$)
the same as those constructed by BMP.

Just as in refs. \WIT,
\REF\BNSR{P. Bouwknegt, J. McCarthy and K. Pilch, CERN
preprint CERN-TH.6346/91 (November 1991).} \BNSR, we may
now construct fields $\Psi_{r,r}\ (r<0)$ forming a
(chiral) ground ring\foot{
We only consider the chiral ground ring here. Extending
the discussion to the full non-chiral ground ring along
the lines of refs. \WIT\ and
\WZ\ is straightforward.}
 and currents $J_{r,r}$ acting on
the ground ring. One finds that the (chiral) ground ring is simply
isomorphic to the ring of non-negative integers (with the
addition) $$\Psi_{r,r} \Psi_{r',r'} = \Psi_{r+r'+1,r+r'+1}
\eqn\cc$$
(modulo BRST commutators),
and that
the currents act trivially:
$$J_{r,r}\left( \Psi_{r',r'}\right) = 0\ .
\eqn\vii$$

Probably the easiest way to see that this must be true is
to remember\foot{
We thank E. Witten for reminding us of this fact.} that we
may perform a rotation (with real coefficients) in the
space of the two fields $\phi_M$ and $\phi_L$ to obtain
new fields $\tilde\phi_M$ and $\tilde\phi_L$ such that
$\tilde\phi_M$ is a $c=1$ (matter) field and
$\tilde\phi_L$ a $c=25$ (Liouville) field. Then eq. \i\
implies $\tilde p_M = {1\over \sqrt{2}} (s-r)$. From our
discussion above we know that hermiticity of the $\cm>1$
matter stress tensor requires $r=s$, which we now
interpret as $\tilde p_M=0$. In other words, the
cohomology of hermitian matter with $1<\cm<25$ is
isomorphic to the one of the $\tilde p_M=0$ sector of
the  $\cm=1$ theory. Of course, we can rotate back
the results of Witten  on the ground ring and
symmetry currents [\WIT] after restriction to this
$\tilde p_M=0$ ($r=s$) sector. The only elements of the
ground ring $x^ny^m$ (in the notation of ref. \WIT ) that
survive the projection are those with $n=m$, and the
projected ground ring is indeed isomorphic to the ring of
non-negative integers. The chiral symmetry currents of the
$\cm=1$ theory act as ${\d h(x,y)\over \d x}{\d\over \d y}
-{\d h(x,y)\over \d y}{\d\over \d x}$, but this can never
map $(xy)^n$ to $(xy)^m$. Hence the symmetry currents that
survive the projection must all be represented by zero
when acting on the projected ground ring. Of course, we
may check this directly in the original $\cm>1$ theory:
$J_{r,r}$ and $\Psi_{r,r}$ carry momenta $\pl=\ql (r+1),\
p_M=\qm (r+1)$. Thus e.g., a priori, $J_{0,0}\left(
\Psi_{-2,-2}\right)$ must be proportional to the identity
$\Psi_{-1,-1}$. Computing the OPE explicitly we see indeed
that the various contributions (proportional to the
identity) cancel among each other and we obtain zero.

To conclude, hermiticity restricts us to $r=s$. The chiral
ground ring is isomorphic to $(xy)^n,\ n=-(r+1)\ge 0$,
while all chiral symmetry currents $J_{r',r'}$ act
trivially ($=0$) on the ground ring.

{ \chapter{ Matter is $D$ free fields: $\cm=D$ (case b).}}

Consider now $D$ free fields (without background charge)
with euclidean or minkowskian signature. As we will see,
the signature is without influence on the result of the
analysis. Although the results can be quite easily be
foreseen from the vanishing theorem of ref. \FGZ, we wish
to stay within the framework of Fock spaces as far as
possible.

We will label the (real) matter momentum by $p$ with
components $p_i,\ i=1,\ldots, D$ and assume first that
$p\ne 0$.
Our strategy is to show that we can always
chose coordinates such that the BMP analysis becomes
particularly simple. In particular we may assume that
$p^+={1\over\sqrt{2}}(p_1+ip_2)$ is non vanishing. We
further define $a_n^\pm={1\over\sqrt{2}}(a_n^1 \pm i
a_n^2)$.

Recall that non-trivial cohomology is only possible
on-shell, i.e. for zero eigenvalue of $L_0=\{ b_0,d\}$,
$b_0$ being the anti-ghost zero-mode and $d$ the BRST
operator\foot{
Indeed, if $L_0 \psi = h \psi$ and $d\psi=0$ then for $h\ne
0$ we have $\psi = (1/h)L_0\psi=d(b_0\psi/h)$.}. The
$L_0=0$ condition reads
$$\half (\pl-\ql)^2-\half\ql^2+\half p^2+R-1=0
\eqn\x$$
where $R$ is the total level operator (matter plus
Liouville plus ghosts) and
$$\ql^2=-{25-D\over 12}\ .
\eqn\xi$$
\def\dh{{\hat d}}
We will consider the relative cohomology which is the
cohomology on the subspace of the Fock space annihilated by
$b_0$ (and $L_0$). On this subspace, the BRST operator
$d=c_0L_0-b_0M+\dh$ reduces to $\dh$ (note $\dh^2=0$).
Following BMP, we break up $\dh$ into three pieces,
$\dh=\dh_0+\dh_1+\dh_2$, of definite degrees 0, 1 and 2
where we define the degree by $$\eqalign{{\rm
deg}(a^+_n)&={\rm deg}(c_n)=1\cr  {\rm deg}(a^-_n)&={\rm
deg}(b_n)=-1\quad (n\ne 0)}
\eqn\xii$$
and assign degree zero to all other operators and to the
Fock space vacua. The benefit of all this is that
$$\dh_0=p^+ \sum_{n\ne 0} c_{-n} a_n^-
\eqn\xiii$$
is particularly simple, and that in the present case it
turns out to be enough to investigate the cohomology of
$\dh_0$. Indeed, one has the following theorem (BMP)
[\BMP] ($C_k$ denotes the subspace of degree $k$):

Theorem: If for each ghost-number $n$ the cohomology
classes $H^n(C_k,\dh_0)$ are non-zero for at most one
degree $k$, then $H^n(C_k,\dh_0)$ and $H^n(C_k,\dh)$ are
isomorphic.

Now, by the usual trick, the cohomology of $\dh_0$ is very
easily obtained by observing that the operator
$K={1\over p^+} \sum_{n\ne 0} b_n a_{-n}^+$ satisfies the
following anticommutation relation with $\dh_0$:
$$\{\dh_0,K\}=\sum_{n\ne 0}\left( a_{-n}^+ a_n^- + nc_{-n}
b_n \right) \equiv \hat R\ .
\eqn\xiv$$
By precisely the same argument as for $L_0$ we now conclude
that non-trivial $\dh_0$-cohomology can only occur for zero
eigenvalue of $\hat R$. Thus the cohomology states are
 all vectors in the Fock space that contain neither
ghost nor  $a^\pm$ excitations. These are ghost number
zero states made up from an arbitrary number of Liouville
and $a^i, i=3, \ldots D$ oscillators such that the on-shell
condition \x\ is satisfied. Since all these states have
zero degree, the above-cited theorem does apply and the
$\dh_0$ and $\dh$ cohomologies are isomorphic. For $R$ a
non-negative integer, let  ${\cal
P}_{D-1}(R)$ denote the number of states at level $R$ in a
Fock space obtained by applying $D-1$ sets of oscillators
$a_{-n}$ to the vacuum. (This is the coefficient of $q^R$ in
$\prod_{n=1}^\infty (1-q^n)^{1-D}$.) Let further
${\cal
P}_{D-1}(R)$ equal zero whenever $R$ is not a non-negative
integer.
 Then we may
state the result for the $\dh$ cohomology - which is in fact
the relative $d$ cohomology - as
$${\rm dim}\, H^n_{\rm
rel}(\fl \otimes {\cal F}(p), d) =\delta_{n0}\, {\cal
P}_{D-1}(R) \quad (p\ne 0)
\eqn\xv$$
where $R=R(\pl,p)$ is determined from the $L_0=0$ condition
\x. Note that the $L_n$ and hence $d$ are rotation/Lorentz
invariant in $D-1$ dimension and that the result \xv, as
well as the $d$ cohomology states, are, of course,
independent of our special choice of coordinates.

It remains to consider the case  $ p=0$.
Then the above construction of $\dh_0$ and $K$
breaks down. We may now define
$p^+(n)={1\over\sqrt{2}}(p_1+i(\pl-(n+1)\ql))=
{i\over\sqrt{2}}(\pl-(n+1)\ql)$ and
$a_n^\pm={1\over\sqrt{2}}(a_n^1 \pm i a_n^L)$ and assign
degrees again by eq. \xii. Then
$$\dh_0=\sum_{n\ne 0} p^+(n) c_{-n}a_n^-\ .
\eqn\xvi$$
If $p^+(n)$ is non-vanishing for all non-zero integers $n$,
i.e. if $\pl \ne \ql(r+1), r\in {\bf Z} (\ne 0)$,
we may construct the corresponding $K$ operator
($K= \sum_{n\ne 0} {1\over p^+(n)} b_n a_{-n}^+$) and
conclude as above to obtain eq. \xv.

However, for $\pl=\ql(r+1)$, we have to exclude the term
$n=r$ from $K$ and the $\dh_0$ cohomology states now may
contain $b_r$ or $a_r^-$ (for $r<0$) and $c_{-r}$ or
$a_{-r}^+$ (for $r>0$) [\BMP] in addition to the $a^i_{-m},
i=2, \ldots D$ oscillators. Thus, in general, the $\dh_0$
cohomology is not confined to a single degree and the
above theorem cannot be applied; we have to obtain the
$\dh$ cohomology in a different way. This will be the only
case where we will go beyond the pure Fock space approach
and use the vanishing theorem of ref. \FGZ.

Note first that for $p=0,\ \pl=\ql(r+1)$ the $L_0=0$
condition becomes
$${25-D\over 24}(r^2-1)=R-1\ .
\eqn\xvii$$
Thus, for given $D$ and $r$, there is not always an
integer solution for the level $R$ where non-trivial
cohomology may occur, If there isn't we just have no
non-trivial cohomology at all for the corresponding
$\pl$. We see also from eq. \xvii\ that (for $D\ne 1$) we
always have $r^2>R$, except for $r^2=R=1$. In the latter
case $\pl=0$ or $2\ql$ and we know that the Liouville and
matter Verma modules have null vectors $L_{-1}^L|p>$
and $L_{-1}^M|p>$ at level $r^2=1$ that vanish in the
Fock space (or in the dual Fock space). So the level of the
(vanishing) null vectors coincides with the level
selected by the on-shell condition \xvii\ and the
vanishing theorem of ref. \FGZ\ does no longer apply ($V$
is not a free ${\cal L}_-$-module). New cohomology states
with non-zero ghost-numbers are possible. Explicitly we
find the cohomology states
$$\eqalign{&a^i_{-1}|\pl=\ql(1+r),p=0> \quad r=\pm 1\
,\cr
&b_{-1}|\pl=0,p=0> \quad (r=- 1)\ ,\cr
&c_{-1}|\pl=2\ql,p=0> \quad (r=+ 1)\ .\cr}
\eqn\xviii$$
We see that with respect to the case of non-zero matter
momentum or $\pl\ne \ql(1+r)$ we find an extra
ghost-number zero and a ghost-number $r(=\pm 1)$
cohomology state.

For $r^2>1$
(and $1<D<25$),
\foot{
Note that there is no possible integer solution  $R$
for $r=0$ and the range of $D$ we are considering.}
the level $R$ selected by the $L_0=0$
condition \xvii\ is always smaller than the level $r^2$
at which the Liouville null vector occurs.
Thus, up to level $R<r^2$ there are no Liouville null
vectors. This is enough to conclude that the combined
Liouville plus matter system has no null vectors up
to level $R$ except when dealing with a state without
Liouville excitations. Then we must still
check for matter null states. In the matter
sector, we have a representation of the Virasoro algebra
with $1<\cm<25$ and highest weight $h=0$. We know from the
Kac determinant formula that, in this range of $c$,
$h_{r's'}$ (with $r's'>0$ since otherwise there are no
null states) equals zero only for $r'=s'=\pm 1$, and the
(matter) null vector is just the obvious
$L^M_{-1}|p=0>$. The latter case corresponds to the
extra cohomology states at level one discussed
above (since these states are on shell only for
$\pl=0, 2\ql$). Thus there are no  null
vectors of the combined Liouville plus matter system for
$p=0$ and $\pl=\ql(r+1)$ with $r\ne \pm 1$ at any level
less than $r^2$.

The BRST operator $d$ does not change the (total) level $R$
of a state. In particular, if we have a state $\psi$ of
level $R$, then the determination of whether $d\psi=0$ or
whether $\psi=d\chi$ only involves elements of the
(matter plus Liouville) Verma module of levels $R'$ with
$R'\le R$. Now, since there are no null vectors in the
(matter plus Liouville) Verma module at any level
$R'\le R<r^2$, we conclude that we may apply  the vanishing
theorem of ref. \FGZ.
\foot{Originally this vanishing theorem has
been proven for a free ${\cal L}_-$-module (${\cal L}_-$
being the negatively moded part of the Virasoro algebra).
Our preceeding argument just shows that it is
sufficient for the cohomology at level $R$ if there are no
null vectors up to level $R$.}
 As a result, for $r\ne\pm 1$ there is
no non-trivial cohomology at non-zero ghost-number, and
the dimensions of the zero ghost-number cohomology groups
are again given by equation \xv\ where $R\equiv R(r)$ is
given by the on-shell condition \xvii.

Finally, one might wish to consider the more general
situation of $D$ free matter fields one of which
(called $\phi_M$) has a background charge $\qm$ (real or
imaginary). By a rotation in field space we may always
reduce this to the case studied above with $\tilde\ql = i
\left( {25-D\over 12}\right)^{1/2}$ as before. Thus the
cohomology is again described by eq. \xv\ except for the
appearance of extra cohomology states (like \xviii) at
level one when $\tilde p=0$ and $\tilde\pl =\tilde\ql
(1+r)$ with $r=\pm 1$, which translates into
$\pl=\ql(1+r)$ and $p_M=\qm(1+r)$ ($r=\pm 1$), and all
other $p_i=0$, in terms of the original (unrotated)
momenta.

\ack

It is a pleasure to acknowledge stimulating discussions
with P. Bouwknegt and E. Witten.

\refout

\end